# *Polarized fine structure in the excitation spectrum of a negatively charged quantum dot*

M.E. Ware[1], E.A. Stinaff[1], D. Gammon[1], M.F. Doty [1], A.S. Bracker[1], D. Gershoni[2], V.L. Korenev[3], Ş. C. Bădescu [1], Y. Lyanda-Geller[4], T.L. Reinecke[1]

[1]*Naval Research Laboratory, Washington DC 20375, USA*

[2]*Physics Department, Technion-Israel Institute of Technology, Haifa 32000, Israel*

[3]*A.F. Ioffe Physical Technical Institute, St. Petersburg 194021, Russia*

[4]*Department of Physics, Purdue University, West Lafayette, IN 47907*

Abstract

We report polarized photoluminescence excitation spectroscopy of the negative trion in single charge tunable InAs/GaAs quantum dots. The spectrum exhibits a *p*-shell resonance with polarized fine structure arising from the direct excitation of the electron spin triplet states. The energy splitting arises from the axially symmetric electron-hole exchange interaction. The magnitude and sign of the polarization are understood from the spin character of the triplet states and a small amount of quantum dot asymmetry, which mixes the wavefunctions through asymmetric e-e and e-h exchange interactions.

PACS No. 78.67.Hc, 71.35.Pq, 71.70.Gm, 72.25.Fe, 71.35.Cc

The spin state of a singly charged quantum dot (QD) could act as a bit of quantum information that is controlled and measured optically through the charged exciton (trion) [1]. Because the optical control of spin information depends profoundly on fine structure, selection rules, and spin interactions, there is considerable interest in high resolution spectral studies of the trion [2]. The trion has been investigated using *non-resonant* excitation in single QDs [3, 4, 5, 6, 7, 8], and in particular, there have been many studies of the 'singlet' trion, in which the two electrons are spin paired. One result of these studies is the demonstration that electron-hole exchange, which plays such an important role in the neutral exciton [9, 10], is absent in the singlet trion [6, 8]. However, in the trion's excited state, in which one of the electrons occupies the *p*-shell, the two electrons can also form a spin triplet. In this case electron-electron (*e-e*) and electron-hole (*e-h*) interactions partly remove the high (8-fold) spin degeneracy, leading to fine structure in the excited state of the trion. Such structure due to spin in quantum dots has been discussed theoretically [11] and invoked to explain non-resonant negative photoluminescence (PL) polarization in InAs dots [12], quantum beating in InP dots at zero field [13], and splittings in multi-exciton and multi-charge PL lines in CdSe [3] and InAs [14] dots.

Here we measure *resonantly* the optical spectrum of a single electron in individual charge-tunable InAs/GaAs QDs for the first time using polarized PL excitation (PLE) spectroscopy. In this spectrum we discover a well-resolved doublet structure with a remarkable polarization reversal that we will assign to trion triplet states.

The InAs QDs were grown by molecular beam epitaxy using an indium flush technique [4, 15]. The QDs were embedded in a Schottky diode to control their electronic charging [5, 7]. The diode structure was grown on a (001) *n*-type GaAs substrate covered with a 500 nm thick GaAs buffer layer (Te doped at ~$5*10^{17}$ $cm^{-3}$), followed by 80 nm of GaAs, the InAs QDs, 230 nm

GaAs, 40 nm $Al_{0.3}Ga_{0.7}As$, and a 10 nm thick GaAs cap. The PL was excited and detected at ~10K through aluminum shadow masks with sub-micron apertures using a *cw*, Ti-Sapphire laser (<7 μeV linewidth) and a 0.5 m spectrometer with a CCD. With voltage controlled retarders for polarization in a strictly backscattering geometry we obtained polarization fidelity greater than 95% at the single dot level.

The charge state of individual QDs was controlled by adjusting the bias across the diode while monitoring the PL spectra. Typical bias dependent PL spectra for nonresonant excitation, $E_{exc}$=2.3 eV, are displayed in Fig. 1a. The image shows the PL emission from both the *s*- and *p*-shells with ~50 meV energy separation between them. Electronic charging in discrete steps is clearly identifiable in the higher resolution *s*-shell spectra in Fig. 1b with excitation energy at $E_{exc}$=1.44 eV, just below the wetting layer energy. The bias at which the dot is charged with a single electron is identified by the appearance of the negative trion, $X^-$ [4, 14].

Low power PLE spectra in which the excitation energy is scanned through the energy range of the *p*-shell emission reveal intense resonances in the $X^-$ with strong polarization dependence (Fig. 2). It is seen (Dot B) that the PLE spectrum of the $X^-$ is shifted down in energy relative to the $X^o$ and becomes strongly polarized when the dot is charged with a single electron. A future analysis of all features will require concepts suggested previously to account for the complex PLE spectrum of a neutral exciton [16, 17], including strong phonon coupling [18], excited state mixing [19], forbidden transitions to the continuum [20, 16, 21], and splitting of the *p*-shells.

In general, we find two very distinctive features. The first is a set of lower energy resonances that are strongly co-polarized with the laser. In Dot B this occurs at ~1302 meV, which is 42 meV above the PL emission. We tentatively assign this structure to nominally forbidden transitions involving states with two *s*-shell electrons and an excited hole [22]. The second feature is a higher energy resonance (for Dot B at ~1313 meV or ~53 meV above the PL) that displays a fine structure doublet well resolved in circular polarization. To date, we have investigated 9 QDs, all of which display this resonant doublet in the range between ~50-57 meV above the $X^-$ PL energy with a splitting of ~216-237 μeV. We will develop an understanding of this doublet in what follows.

High-resolution PLE spectra of the doublet for the two circular PL polarizations are shown in the insets of Fig. 2. For all dots, both lines exhibit extremely large polarizations ranging from ~65% to greater than 95% calculated by $P=(I^+-I^-)/(I^++I^-)$, with $I^+$ ($I^-$) representing the integrated intensity of the $\sigma^+$ ($\sigma^-$) polarized PLE resonance under $\sigma^+$ polarized excitation. The halfwidths of each of the lines are ~50 μeV in the low power limit. In all the QDs the low-energy component is positively polarized, and the high-energy line is negatively polarized, i.e., parallel and anti-parallel with the laser, respectively. In the following we argue that this PLE resonance and its polarized fine structure result from direct excitation into the *triplet states* of the excited trion (see Fig. 2; left inset.)

In Fig. 3 we consider in more detail the optical excitation of an *e-h* pair directly into the *p*-shell in the presence of an unpolarized "resident" *s*-shell electron [11]. The electrons will form symmetric spin triplets and an anti-symmetric spin singlet split by the *e-e* exchange of $\Delta_{ee} \approx 6\text{ meV}$ [5]. The triplet states are further split by the axially symmetric part of the *e-h* exchange interaction ($\Delta_{eh}$) into a set of three degenerate doublets with the order given in Fig. 3. This splitting is expected to be several hundred μeV as deduced from the transverse magnetic field dependence of the $X^o$ [6], and from the fine structure of the $X^{2-}$ [14] and $2X^-$ lines [14, 3]. In Fig. 3 the trion states have been sorted into two columns that show vertical optical transitions under circularly polarized excitation. Only four of the six triplet states are optically allowed.

Thus, resonant $\sigma^+$ light should lead to excitation of two states split by $\Delta_{eh}$ (wavy 'up' arrows in Fig. 3), as observed. These two transitions differ in the initial state by the spin projection of the resident electron, and thus the laser frequency and polarization provide a spin selective probe.

To understand the dramatic polarization reversal in the fine structure doublet we consider relaxation from the excited trion *triplet* states (populated by $\sigma^+$-light) to the luminescing *s*-shell *singlet*. The relaxation of the low-energy $m_s = +3/2$ triplet state (left side of Fig. 3) conserves both the hole spin and the momentum projection of the electrons. The low energy resonance should therefore retain its $\sigma^+$ character during relaxation and recombination, as observed. The $m_s = +1/2$ high energy state of the excited trion triplet (right side of Fig. 3), however, requires a change of momentum projection of -2 in order to relax through a projection conserving channel. It is well understood that single particle spin flips are strongly suppressed in QDs [23]. However, the asymmetric part of the *e-h* exchange interaction ($\delta_{eh}$) has been shown to mix the $m_s = \pm 1/2$ states with the $m_s = \mp 3/2$ states by a small amount [11]. This simultaneously flips the *p*-shell electron and the hole spins leading to relaxation to the ground state singlet from the $m_s = \pm 1/2$, excited states [12]. (This is necessary, but not sufficient as will be discussed below.) With the hole now having opposite spin, the emission is thus polarized anti-parallel to the laser light ($\sigma^-$ emission).

The absolute magnitudes of the polarizations allow us to deduce the amount of wavefunction mixing ($\alpha$), which in turn determines $\delta_{eh}$ from the relation $\alpha = (\delta_{eh}/\Delta_{eh})^2$ [11] with $\Delta_{eh}$ given by the splitting of the doublet. Because the oscillator strengths of the transitions differ, the absolute polarizations are different for each channel. The oscillator strength of the $\sigma^+$-polarized optical transition into the state denoted by $|m_s = +3/2\rangle = |\Uparrow\rangle(|\uparrow\downarrow\rangle + |\downarrow\uparrow\rangle)/\sqrt{2}$ from the $|\uparrow\rangle$ electron state is 2 times smaller than that for excitation into $|m_s = +1/2\rangle = |\Uparrow\rangle|\downarrow\downarrow\rangle$ from $|\downarrow\rangle$. We find polarization values of ~70% for $\sigma^+$ and ~-90% for $\sigma^-$ PLE lines which gives a value of $\alpha \approx 0.08$, corresponding to $\delta_{eh} \approx 60\ \mu\text{eV}$, which is consistent with our measured neutral exciton fine structure splittings in PL ($\delta_{eh} \lesssim 40\ \mu\text{eV}$) [9, 6]. Thus, the model provides a natural explanation of the magnitude of the fine structure splitting, the signs and values of the strong polarization, and the order of PLE resonances observed in the resonant *p*-shell PLE spectra of the negative trion.

Interestingly, the two PLE components are power-broadened [24], but at substantially different rates, as shown in Fig. 4a and b. We analyze these data to obtain estimates of the relative relaxation rates. Power broadening arises from the saturation of the transition, thus the $\sigma^-$ PLE line, with its larger oscillator strength, broadens faster than the $\sigma^+$ line. The integrated area of each of the PLE lines (Fig. 4c) reveals a similar trend. Fits to the data in Fig. 4b and c are obtained using density matrix equations and a ratio of 2 for the oscillator strengths, thus allowing us to estimate the triplet to ground state singlet relaxation rates as $\gamma_+ \approx (25\ \text{ps})^{-1}$ and $\gamma_- \approx (310\ \text{ps})^{-1}$ from the $m_s$=3/2 and $m_s$=1/2 states, respectively. As we expect, due to the asymmetric *e-h* exchange mixing, the $m_s$=1/2 state relaxes an order of magnitude slower than the $m_s$=3/2 state. The PLE intensities are comparable, because the rate limiting process is the PL radiative rate, $\Gamma_R \approx (1\ \text{ns})^{-1}$, which is the same for both [12].

As discussed above, asymmetric *e-h* exchange mixes the $m_s = \pm 1/2$ with the $m_s = \mp 3/2$ triplet states, conserving the spin projection in relaxing to the ground state singlet, and explaining the unique PL polarization we have observed. However, this relaxation involves a change in the *total* electron spin of one and should be forbidden. To resolve this contradiction, we also consider a mixing of the $m_s = \pm 3/2$ triplet states with the *excited* singlet states, now through the asymmetric part of the *e-e* exchange (Fig. 3) [25]. The Dzyaloshinskii-Moriya form of this exchange, given by $\vec{\delta}_{ee} \cdot \vec{s}_1 \times \vec{s}_2$, arises from spin orbit coupling [26, 27] and can mix the triplet states of electrons with the corresponding *excited* singlet states, which relax quickly to the ground state singlets (within $\tau \approx 1\,\text{ps}$). Using $\gamma_+ \approx (\delta_{ee}/\Delta_{ee})^2 \tau^{-1}$ to estimate the relaxation rate of the triplets, our results indicate that $\delta_{ee}/\Delta_{ee} \approx 1/5$. We have obtained the asymmetric exchange for a single dot within a Kane model calculation [27] and find a term $\delta_{ee}(\vec{s}_1 \times \vec{s}_2)_z$ that mixes the excited singlet and the triplet states if the potential has inversion asymmetries in the lateral directions. We have evaluated $\delta_{ee}$ for dots with modest cubic and fourth order terms added to the lateral parabolic potentials. The value $\delta_{ee}/\Delta_{ee} \approx 1/5$ is obtained with an anharmonicity of $\lesssim 12\%$ as seen in the spacing of the excited states. We believe this deviation from an ideal potential to be reasonable [28] and thus conclude that asymmetric *e-e* exchange can lead to fast relaxation from triplet to ground state singlet.

Recently, negative PL polarization of the trion with *non-resonant* circularly polarized excitation was explained with two alternative models [12, 29, 30]. The first [29, 30] is based on the creation and accumulation of dark excitons due to single spin flips of delocalized holes. Subsequent capture by a negatively charged QD and recombination leads to negative polarization. The second model [12] is based on the mutual *e-h* flip-flop transition mediated by the asymmetric *e-h* exchange in the triplet state. Recently, we proved [30] that the first mechanism dominates under non-resonant excitation in GaAs/AlGaAs QDs. Now, using polarized *resonant* excitation of the triplet states in InAs/GaAs QDs, where single carrier spin flips are suppressed, we have demonstrated directly the second mechanism for negative PL polarization.

In conclusion, we have discovered a fine structure doublet in the polarized PLE spectra of the negative trion. We show that the doublet arises from the resonant excitation of triplet states of the trion with splitting due to the symmetric *e-h* exchange interaction. Furthermore, the asymmetric part of the *e-h* exchange mixes the triplet states, while the asymmetric part of the *e-e* exchange mixes the excited singlet with the triple states. This leads to two separate relaxation channels that are selected by laser energy and recombine with both positive and negative polarization. Finally, we note that because the components of the doublet arise from opposite spin states of the resident electron (Fig. 3), these polarized resonant transitions provide an all optical PLE method of selectively probing the spin state of a single electron confined in a QD.

We acknowledge the financial support of DARPA/QuIST, ARDA/NSA/ARO, ONR, CRDF, RFBR, RSSF. MW, ES, MD, and SB hold NRC/NRL Associateship Awards.

Figure Captions

Fig. 1 Bias dependent PL intensity maps. (a) Emission from both the *s*- and *p*-shells of a single dot. (b) Charging structure of the *s*-shell emission from another dot.

Fig. 2 Polarized PLE spectra of the $X^-$ under $\sigma^+$ excitation for two dots. The PL emission energies are given in parenthesis. The top (bottom) trace for each spectrum is for $\sigma^+$ ($\sigma^-$) PL detection. The polarized doublet for each dot is expanded in the insets. For Dot B, the associated $X^o$ PLE spectrum is also displayed. The left inset shows the energy levels of the ground ($X^-_{ss}$) and excited ($X^-_{sp}$) state trions with and without *e-e* ($\Delta_{ee}$) and *e-h* ($\Delta_{eh}$) exchange splittings.

Fig. 3 Energy levels of the negative trion states. The $\Uparrow$ ($\uparrow$) arrowheads represent the spin projection of the holes (electrons) with $m_s$ the total spin projection of each state. The excited singlet is split from the triplet by $\Delta_{ee}$ and the excited triplets are split by $\Delta_{eh}$; the symmetric exchange energies. The mixed states are indicated by the dashed arrows, with $\delta_{ee}$ and $\delta_{eh}$ the asymmetric exchange energies which allow the mixing.

Fig. 4 (a) Power dependence of the polarized doublet. (b) PLE halfwidth. The solid lines are fits to $\Delta_\pm(P) = \hbar\Gamma_{d\pm}\sqrt{1+b_\pm P}$. (c) Integrated area of each PLE line with fit to $I_{Area}(P) = a_\pm P \big/ \sqrt{1+b_\pm P}$. Parameters $a_\pm$ are proportional to the oscillator strengths of the $\sigma^\pm$-PLE transitions ($a_- / a_+ = 2$) whereas parameters $b_\pm$ also include the relaxation rates to the s-shell and recombination rates. The fitted parameters are as follows: $a_+ = 63.5\,\text{mW}^{-1}$, $b_+ = 0.472\,\text{mW}^{-1}$, $\hbar\Gamma_{d+} = 48.7\,\mu\text{eV}$, $a_- = 2a_+$, $b_- = 1.68\,\text{mW}^{-1}$, $\hbar\Gamma_{d-} = 37.1\,\mu\text{eV}$.

Ware, et al., Figure 1

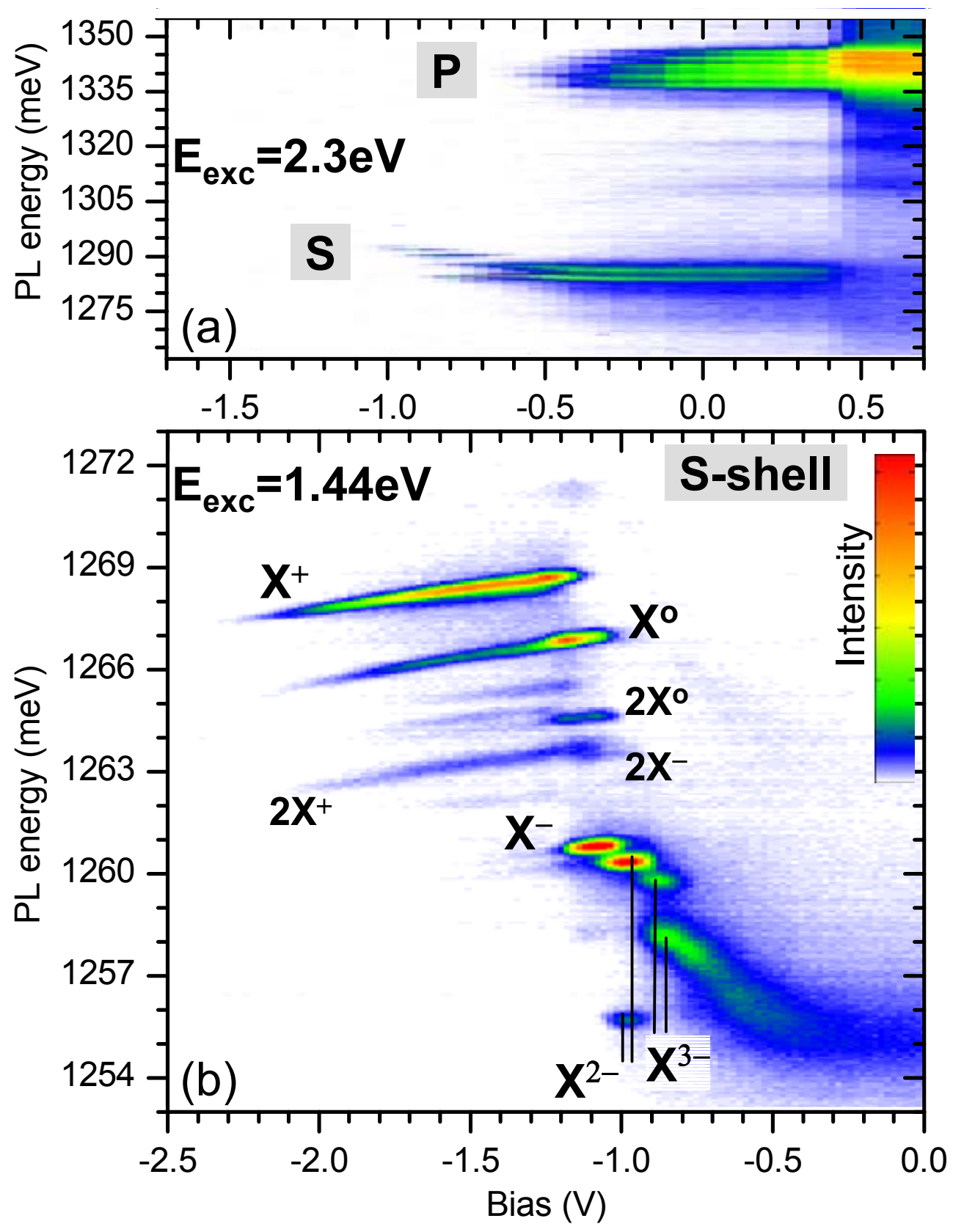

Bias dependent PL intensity maps. (a) Emission from both the *s*- and *p*-shells of a single dot. (b) Charging structure of the *s*-shell emission from another dot.

Ware, et al., Figure 2

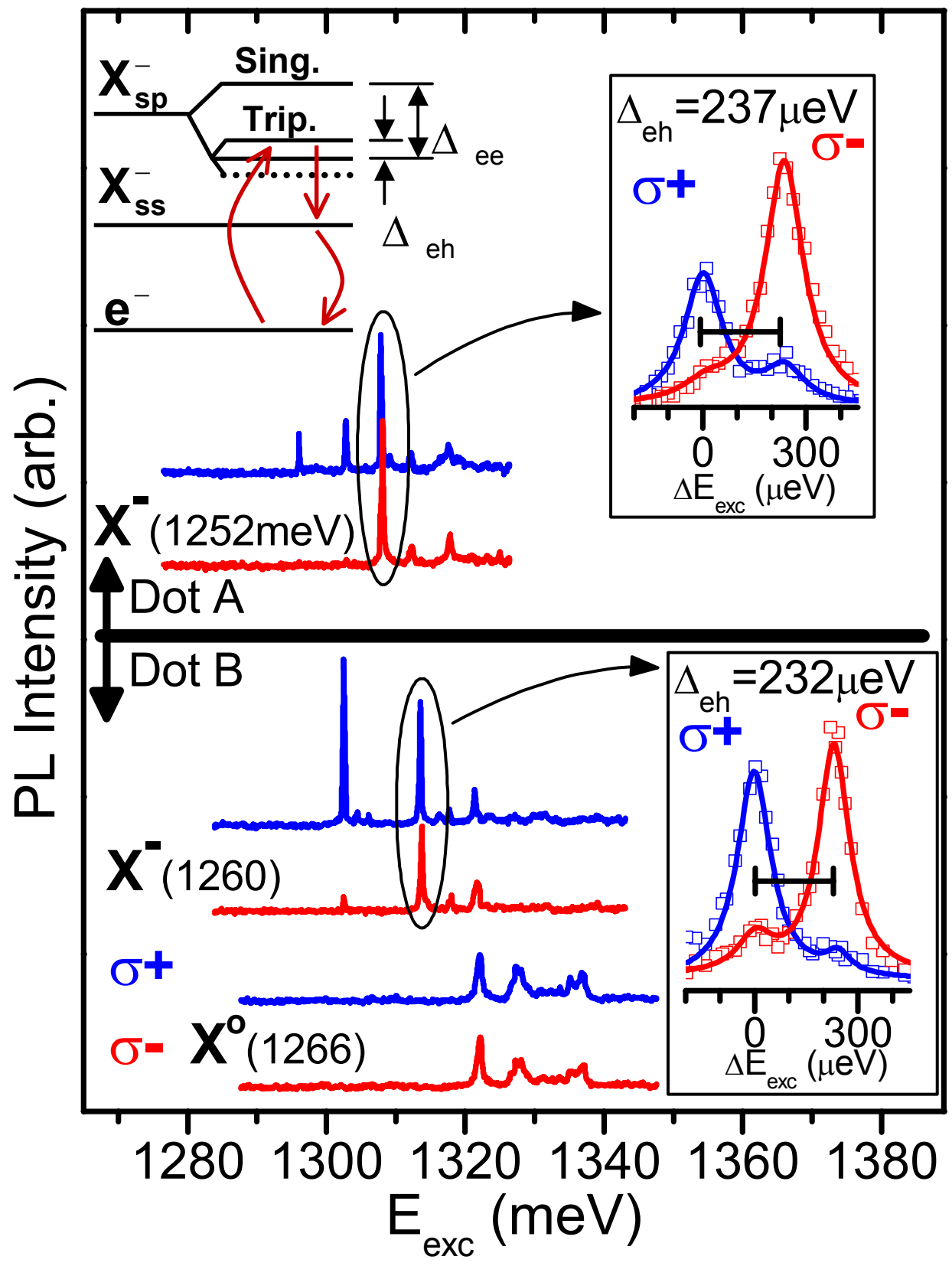

Polarized PLE spectra of the $X^-$ under $\sigma^+$ excitation for two dots. The PL emission energies are given in parenthesis. The top (bottom) trace for each spectrum is for $\sigma^+$ ($\sigma^-$) PL detection. The polarized doublet for each dot is expanded in the insets. For Dot B, the associated $X^o$ PLE spectrum is also displayed. The left inset shows the energy levels of the ground ($X^-_{ss}$) and excited ($X^-_{sp}$) state trions with and without *e-e* ($\Delta_{ee}$) and *e-h* ($\Delta_{eh}$) exchange splittings.

# Ware, et al., Figure 3

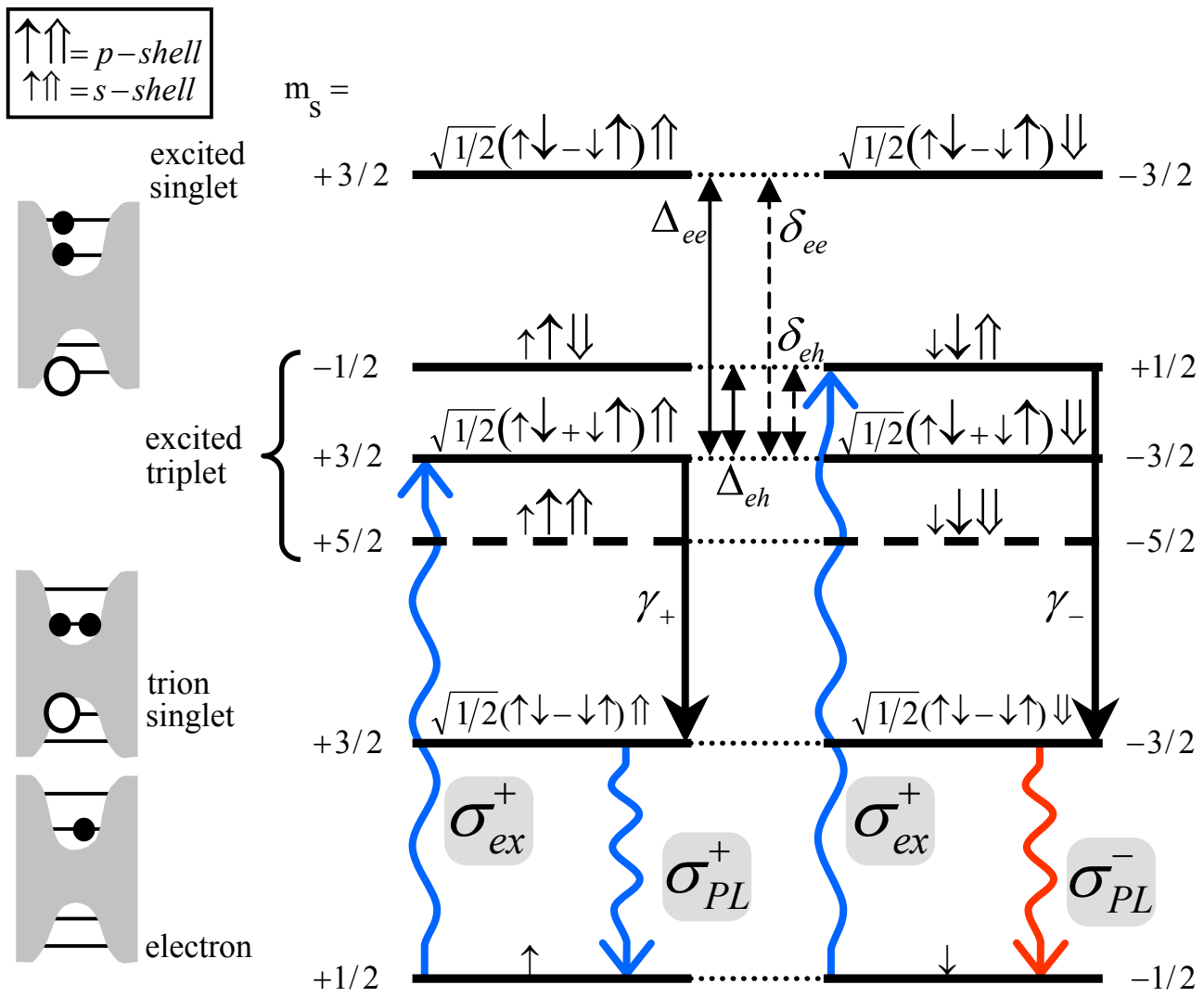

Energy levels of the negative trion states. The ⇑ (↑) arrowheads represent the spin projection of the holes (electrons) with $m_s$ the total spin projection of each state. The excited singlet is split from the triplet by $\Delta_{ee}$ and the excited triplets are split by $\Delta_{eh}$; the symmetric exchange energies. The mixed states are indicated by the dashed arrows, with $\delta_{ee}$ and $\delta_{eh}$ the asymmetric exchange energies which allow the mixing.

# Ware, et al., Figure 4

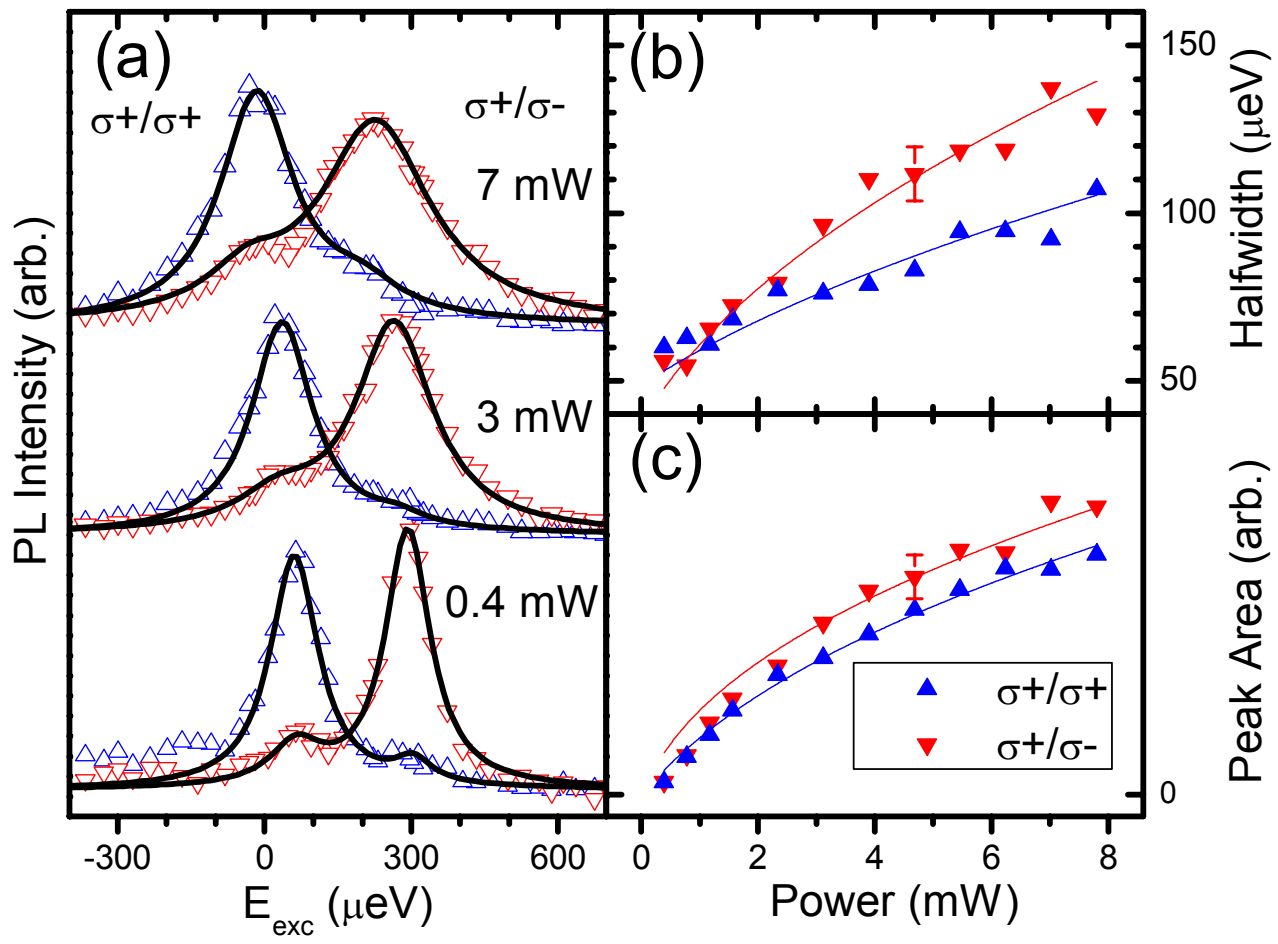

(a) Power dependence of the polarized doublet. (b) PLE halfwidth. The solid lines are fits to $\Delta_{\pm}(P) = \hbar\Gamma_{d\pm}\sqrt{1+b_{\pm}P}$. (c) Integrated area of each PLE line with fit to $I_{Area}(P) = a_{\pm}P/\sqrt{1+b_{\pm}P}$. Parameters $a_{\pm}$ are proportional to the oscillator strengths of the $\sigma^{\pm}$-PLE transitions ($a_{-}/a_{+} = 2$) whereas parameters $b_{\pm}$ also include the relaxation rates to the s-shell and recombination rates. The fitted parameters are as follows: $a_{+} = 63.5\,\text{mW}^{-1}$, $b_{+} = 0.472\,\text{mW}^{-1}$, $\hbar\Gamma_{d+} = 48.7\,\mu\text{eV}$, $a_{-} = 2a_{+}$, $b_{-} = 1.68\,\text{mW}^{-1}$, $\hbar\Gamma_{d-} = 37.1\,\mu\text{eV}$.